\documentclass[preprints,article,accept,pdftex,moreauthors]{Definitions/mdpi} 

\firstpage{1} 
\pubvolume{1}
\issuenum{1}
\articlenumber{0}
\pubyear{2023}
\copyrightyear{2023}
\datereceived{ } 
\daterevised{ } 
\dateaccepted{ } 
\datepublished{ } 
\hreflink{https://doi.org/} 

\Title{Astrochemistry of the molecular gas in  Dusty Star-Forming Galaxies at the Cosmic Noon}

\TitleCitation{Molecular gas in DSFGs}

\Author{Francesca Perrotta $^{1}$\orcidA{}, Martina Torsello$^{1}$\orcidB{}, Marika Giulietti$^{2}$ \orcidC{}, Andrea Lapi $^{1}$ \orcidE{}}

\AuthorNames{Francesca Perrotta}

\AuthorCitation{Perrotta F.}

\address{%
$^{1}$ \quad Scuola Internazionale Superiore di Studi Avanzati, Via Bonomea 265, 34136 Trieste, Italy \\
$^{2}$ \quad INAF - Istituto di Radioastronomia,
Via Gobetti 101, 40129 Bologna, Italy\\
}

\corres{Correspondence: perrotta@sissa.it}
\nolinenumbers
\abstract{Far-infrared and sub-millimeter observations have established the fundamental role of dust-obscured star formation in the assembly of stellar mass over the past $\sim 12$ billion years. At z= $2-4$, the so-called "cosmic noon", the bulk of star formation is enshrouded in dust, and dusty star forming galaxies (DSFGs) contain $\sim 50 \%$ of the total stellar mass density. 
Star formation develops in dense molecular clouds, and is regulated by a complex interplay between all the ISM components that contribute to the energy budget of a galaxy: gas, dust, cosmic rays, interstellar electromagnetic fields, gravitational field, dark matter. Molecular gas is the actual link between star forming gas and its complex environment: much of what we know about star formation comes from observation of molecular line emission. They provide  by far the richest amount of information about the star formation process. However, their interpretation requires complex modeling of astrochemical networks, which regulate the molecular formation and establishes molecular abundances in a cloud, and a modeling of the physical conditions of the gas in which molecular energy levels become populated. This paper critically reviews the main astrochemical parameters needed to get predictions about molecular signals in DSFGs. Molecular lines can be very bright compared to the continuum emission, but radiative transfer models are required to properly interpret the observed brightness. We review the current knowledge and the open questions about the interstellar medium of DSFGs, outlying the key role of molecular gas as a tracer and shaper of the star formation process. }


\begin{document}


\section{Introduction}
The first detection of high-redshift dusty, star-forming galaxies (DSFGs) dates back to the SCUBA observations in the late 1990s \cite{Smail1997},\cite{Smail1998},\cite{Blain2002}. It soon became clear that they constitute a cosmologically significant population of galaxies at z=2-4, appearing extremely bright in the sub-mm and IR  ($S_{870 \  \mu m} \ \geq  1$ mJy, with typical L$_{\rm IR} \sim 10^{12}$ L$_\odot$), but nearly invisible in the optical \cite{Hodge2013},\cite{Simpson2015a}, \cite{Simpson2015b}, \cite{DaCunha2015}, \cite{Oteo2016},\cite{Casey2014}, \cite{Greve2012}. The brightest of these galaxies have infrared luminosities exceeding $10^{13} $ L$_{\odot}$. Because of this, DSFGs  were formerly known as Sub-millimeter Galaxies (SMGs).  The impressive  DSFGs infrared luminosity reveal star formation rates (SFRs) as high as a few $ 10^2-10^3 $ M$_{\odot}$ yr$^{-1}$ \cite{Gruppioni2013}, \cite{Bethermin2015}. \\
The  widespread interpretation  is that SMGs are dominated by short-lived and extreme burst events, in which the high rate of star formation is accompanied by the rapid production of large amounts of dust. The dust grains, in the Interstellar Medium (ISM) surrounding active star forming regions are heated mostly by the UV light from massive, young stars, and then cooled down via thermal emission of infrared radiation, producing the typical intense IR brightness observed (redshifted, at high z, to far infrared (FIR) and submm  wavelengths), while obscuring 95$\%$ the stellar emission(see, e.g. \cite{Casey2014} and \cite{Forster2020} for extensive reviews). \\
In the last two decades, huge effort has been devoted to perform higher resolutions and multiband observations of DSFGs, which became available  thanks to the advent of facilities such as the Acatama Large Millimeter Array (ALMA), Northern Extended Millimeter Array (NOEMA), VLA, South Pole Telescope, Spitzer, $Herschel$, and the most recent James Webb Space Telescope. 
Ever deeper and wider look-back  surveys allowed to establish a fairly robust outline of the cosmic history of star formation (SF), which culminated at z$\sim 2-3$  (10 billions year ago, or few Gyrs after the Big Bang: the often nicknamed  "cosmic noon"). At cosmic noon, the bulk of star formation activity is strongly enshrouded in dust (see, e.g., \cite{Casey2014}, \cite{Zavala2021}),  and chiefly dominated by the most luminous and massive DSFGs (with L$_{\rm IR} >  10^{12}$ L$_\odot$),  with DSFGs containing $\sim 50\%$ of the total stellar mass density \cite{Swinbank2014}.  \\
Furthermore, the redshfit evolution of the cosmic SFR density remains dominated by dust-obscured SF at least  over the past $\sim 12 $ billion years, back to z $\sim 4$ \cite{Swinbank2014}, \cite{Pantoni2021a}, \cite{Pantoni2021b}.  \\
Thus, characterizing the star-formation process in the DSFGs population is of paramount importance to understand the stellar mass assembly of the young Universe:  what triggered the fast, intense burst of star formation observed at the "cosmic noon", and which is the mechanism that eventually quenched it? 
Despite two decades of observations and modeling, a unique explanation is not yet available. 
Star formation is the result of the interplay between diverse, though interconnected, complex physical processes occurring in the many phases of the Interstellar medium (ISM) and Intergalactic medium (IGM), eventually culminating in the fragmentation and collapse of molecular clouds (e.g. \cite{Tacconi2020}, \cite{Krumholz2011}, \cite{Kim2021}): with the possible exception of Population III stars, SF is ultimately fueled by molecular gas. Several global-scale and local-scale galactic parameters compete in the  SF process and in shaping the Initial Mass Function (IMF), such as dust, gas and stellar content, gas kinematics, environment and local radiation fields, magnetic fields, global and local cosmic ray flux, metallicity, balance of heating and cooling, infall of cold gas from the IGM (see, eg. \cite{Granato2004}, \cite{Putman2017}, \cite{Pantoni2019}, \cite{Suzuki2021}).  
  Molecular clouds,  potential cradle of stars, are supported by magnetic fields and turbulence against gravitational collapse, but the knowledge of the processes ruling  the clouds lifecycle and their star formation efficiency (defined as the star formation rate per unit of interstellar gas) remains an open question \cite{Chevance2020}. \\
Thus, the ultimate question necessarily involves the molecular component of the ISM (e.g., \cite{Garciaburillo2012},\cite{Rybak2022}): in which respects the molecular gas of local star-forming galaxies (ULIRG and LIRG) is different from that of high-redshift DSFGs? Is star formation described by a universal law which applies to different galaxy populations?
How can the observation of molecular lines help in answering these questions? The answer requires a proper characterization  of the diffuse and dense molecular ISM phases at high redshifts, which is generally observed through rotational transitions of the most abundant "tracers" molecules, whose brightness can still be detected even at high z.     \\
Diagnostics based on molecular line intensities, and on their ratios, goes through the comparison between spectral observations and models of line emission from key molecular species, both  in dense clouds and in the diffuse molecular medium. In turn, this requires an accurate numerical solution of the radiative transfer problem, implying the characterization of a  plausible physical environment  in which the  emitted line radiation propagates. But this is not sufficient yet: line intensities ratios between different molecular species may suffer of a degeneracy between excitation effects and abundances effects, even when they arise from the same physical region in the host galaxy. It is then necessary to provide a solid astrochemical model to predict the relative molecular abundances in a given environment and to compare them with real data. 
Since the discovery of the first interstellar molecules CH, CN, CH$^+$ (\cite{Swings1937}, \cite{McKellar1940}, \cite{Douglas1941}), hundreds of molecules and isotopologues have been detected, including the breakthrough discovery of Complex Organic Molecules throughout the Milky Way and in nearby galaxies, as well as in some distant quasars, which opened new scenarios for the emergence of life in exoplanets (e.g. \cite{Ceccarelli2003},\cite{Omont2007}, \cite{Millar2015}, \cite{Sewilo2019}, \cite{Cernicharo2021},\cite{McGuire2022},\cite{Guelin2022}). To date, several  galaxies have been mapped in the emission lines of diatomic or more complex molecules. 
In particular, in  the last two decades, DSFGs have been extensively observed in their dust continuum emission and in the  CO (\cite{Greve2005}, \cite{Solomon2005}, \cite{Omont2007}, \cite{Carilli2013b}, \cite{Bothwell2013}, \cite{FriasCastillo2022},\cite{Giulietti2023}, \cite{Ivison2010}, 
\cite{Danielson2011},  \cite{Kirkpatrick2019}
\cite{Yang2017} \cite{Gururajan2022}, \cite{Weiss2005}), H$_2$O , neutral Carbon and [CII]  emissions  (\cite{Carilli2013}, \cite{Hodge2020}, \cite{Riechers2011a}, \cite{Riechers2011}, \cite{Dannerbauer2017}, \cite{Rybak2019}, \cite{Birkin2021}, \cite{Rybak2020}, \cite{Perrotta2023} ), while only few HCN detections have been established to date \cite{Rybak2022}, \cite{Yang2023}.  \\
Stacked spectra of DSFGs samples have been obtained by \cite{Spilker2014} (in which the average mm/submm rest-frame spectrum was constructed by stacking the ALMA spectra of 22 gravitationally lensed  sources spanning a redshift range of z = 2.0–5.7 and rest frame-frequencies 250–770 GHz, or $\lambda$= 0.39–1.2 mm) and by \cite{Wilson2017} which built the average far-infrared spectra of a sample of 36 DSFGs at 0.8 < z < 4 
with rest frame  frequencies 1400–6200 GHz ($\lambda$ =48–214 $\mu$m). The analysis of FIR  stacked spectra for the highest resdshift sources allows to obtain general informations about the typical state of the ISM in DSFGs, and it is suggestive of an intense FUV field dominating the diffuse gas (a factor of $10^3-10^5$ higher than the Milky Way FUV field) and of an average density of the neutral gas of about $10^{4.5}-10^{5.5}$ cm$^{-3}$. The stacked submm spectrum revealed, for the first time at those high redshifts, the emission from the hydride CH and the linear molecule CCH, which may turn out to be important probes of  astrochemistry in DSFGs. Furthermore,  the stacking technique allowed to detect emissions from the high-density molecules HCN, HNC, HCO$^+$ and CS, often elusive in observations of single sources at that redshift. \\
In some cases, relatively detailed observations of single sources were made possible thanks to the size stretching and intensity amplification offered by a foreground gravitational lens. \\
In parallel, astrochemistry underwent a major development, pushed by the need to identify the main mechanism driving the molecular enrichment in diverse ISM environments.  The huge effort of the astrochemistry community has lead to the identification of the main chemical reaction networks and their reaction rate coefficients, corresponding to the main physical driver of the reactions and, ultimately, to the physics of the ISM. \\
The first attempts to solve gas phase and grain surface chemical reactions networks on physical environments representative of external galaxies date back to the works of \cite{Bayet2008}, \cite{Bayet2009}. 
In order to decipher the information hidden in spectral lines from high-z DSFGs, the  methodology clearly needs to be reviewed with respect to the local observations. While sticking to the universality of the underlying physical-chemistry, unlensed DSGFs molecular lines represent source-integrated signals, showing up as surface-averaged molecular line fluxes.  Thus, a  drawback of  extrapolating the local astrochemical fundamentals to the extragalactic context, lies in the misleading use of the so called "molecular tracers": while the emission line fluxes of selected molecules, and their ratios, may be locally suggestive of specific ISM physical conditions (such as an high or low  density, aligned with the critical density of the molecular transition line, or  an interstellar radiation field dominated by the extreme emission of an Active Galactic Nuclei), the unresolved emission from a distant galaxy will necessarily mix and integrate molecular emissions from different regions of the galaxy, plausibly pertaining to different ISM phases. 
The unavoidable lower resolution, with respect to nearer, local sources, forces us to put an effort in  simplifying  the problem, e.g.  waiving the detailed analysis of the individual hot cores or corinos, and trying to analyze clusters of forming stars, giant molecular clouds and diffuse molecular components, in a statistical, averaged meaning. The starting point, however, is the characterization of the ISM in which molecular clouds emerge. 

If the evolution of each galaxy is a tale of star formation, than the exploitation of galaxy evolution cannot overlook the physics and chemistry of the ISM. In this respect, galaxy evolution is a tale of ISM, and, ultimately, a tale of how molecules builds out of elements.  \\
 Here we want to provide some considerations to be accounted for when using astrochemical networks to model the molecular abundances in  high-redshift DSFGs. The main question behind is what's peculiar in  the ISM of high-z DSFGs with respect to the local star forming galaxies,  with emphasis on the molecular gas: trivially, they are at higher redshifts than the Milky Way, so the CMB radiation was warmer than today; they have high rates of star formation, which means that the interstellar radiation field and the flux of cosmic rays are stronger than they are in quiescent galaxies, which  plausibly deeply affects the molecular abundances; 
 they harbor large amounts of dust which catalyzes molecular formation on grain surfaces, affecting the gas phase molecular abundances, and regulates, through continuum opacity, the onset of cosmic rays as chemical drivers at large enough optical depths in clouds. \\
We review some of the ISM characteristics in DSFGs at cosmic noon, as they emerge from observations, the influence of the environmental ISM in the molecular abundances, and the open questions that need to be solved in order to get reliable informations from molecular observations in such extreme environments. \\
The structure of the paper is the following: in the first part  (Sec. \ref{sec:cmb}-\ref{sec:dust})
we present an overview of the ISM environment of high-z DSFGs: in Sec. \ref{sec:cmb} we discuss the impact of CMB temperature at cosmic noon on the molecular emission lines; in Sec. \ref{sec:FUV} we characterize the far UV interstellar radiation field; in Sec. \ref{sec:cosmic_rays} we describe the role of cosmic rays as drivers of astrochemistry in the molecular clouds where the FUV field is attenuated, discussing observations and molecular signatures of cosmic rays ionization; in Sec. \ref{sec:dust} we describe the fundamental role of dust grains on the physics and chemistry of DSFGs. The second part of the paper summarizes the state-of-art theory and observations of the principal molecules detected in high-z DSFGs molecular gas: H$_2$ and CO in Sec. \ref{sec:h2co}, water vapor in Sec. \ref{sec:H2O}, dense gas tracers in Sec. \ref{sec:dense}. Finally, in Sec. \ref{sec:summary} we conclude with our final remarks. 

\section{\bf The effect of CMB on the molecular gas of high-z galaxies}
\label{sec:cmb}
The standard hot Big-Bang model predicts the linear increase with redshift of the Cosmic Microwave Background (CMB)  temperature, as a consequence of the adiabatic expansion of the Universe. As CMB photons propagate along null geodesics, the  CMB temperature varies as  T$_{\rm CMB}$= T${^0}_{\rm CMB}$  $\times (1+z)$, where the current value  $ {{\rm T}^0}_{\rm CMB}= 2.725 1\pm 0.002$ K \cite{Fixsen2009}  corresponds to a blackbody spectrum peaking at ${\nu}^0_{\rm max} \sim 2.82 \ k{\rm T}/h \sim 160$ GHz.  At $z=4$, this corresponds to T$_{\rm CMB}$ = 13.625 K and to a blackbody spectrum peaking at $\sim 800 $ GHz.\\
The energy density of blackbody radiation, integrated over all the spectrum, is given by 
$u({\rm T})= a \ {\rm T}^4$, being $a$ the Stefan-Boltzmann constant
$a=7.56 \times 10^{-15}$ erg cm$^{-3}$ K$^{-4}$. \\
The current value of the CMB energy density is $u=4.19 \times 10^{-13}$ erg cm$^{-3}$: in the Milky Way, the CMB dominates the Galactic spectrum above 1 GHz and up to $\sim 500 $ GHz, where thermal emission  from dust at T$_{\rm dust} \sim 18$ K takes over.  Since the CMB energy density scales as T$_{\rm CMB} ^4 \propto (1+z)^4$, in galaxies  at , e.g., z$\sim 3$, it was larger than its current value by a factor $> 200$. Similarly,  the number density of blackbody photons, integrated over all the spectrum, scales as 
\begin{equation}
n_{\gamma}=16 \pi \zeta(3) {\left(   \frac {k{\rm T}} {hc} \right)}^3
\end{equation} 
where $\zeta$ is the Riemann function, which has $\zeta (3) \approx 1.202$. 
From the above scaling relations, we should expect the CMB to have an impact on high-z galaxies, already at cosmic noon. \\
The CMB temperature sets the fundamental minimum temperature of the ISM (assuming local thermal equilibrium) and can affect the physical conditions of dust and gas, in particular the molecular ISM. For instance, a remarkable thermometer of the CMB temperature in the Milky Way is the CN molecule \cite{Meyer1985}, \cite{Kaiser1990}, \cite{Roth1993}. \\
A higher CMB temperature has two competing effects on the physics of molecular clouds: on one hand, the higher dust and gas temperature corresponds to a boosted  luminosity of emission lines and of the dust continuum \cite{Silk1997}, \cite{Blain1999}, \cite{Combes1999}, \cite{Righi2008}. On the other hand, the warmer CMB builds a stronger background against which dust continuum and emission lines are detected \cite{Combes1999}, \cite{Papadopoulos2000}, \cite{Obreschkow2009}, \cite{Lidz2011}, \cite{Munoz2013}: at increasing T$_{\rm CMB}$, the thermal equilibrium between the CMB, the cold gas and the dust, progressively erases the spectral contrast which makes dust and line emissions detectable. \\
A full exploration of the CMB effects on the observability of emission lines and dust continuum is presented in \cite{DaCunha2013}, with emphasis of the Carbon monoxide (CO) lines. Indeed, CO rotational levels have energy differences which are close to kT$_{\rm CMB}$ at high redshift; in general, if the molecular gas is  permeated by a bath of photons whose frequencies are distributed according the Planck law for a blackbody at temperatures of about 13 K, there will be a large number of photons in the Rayleigh-Jeans tail of the distribution that may pump rotational levels of CO, H$_2$O and other molecules frequently detected in the high-z DSFGs.  \cite{DaCunha2013} found that the dominant effect is that of attenuating the observed line and continuum flux because of the enhanced brightness temperature of the background.  Neglecting the influence of the CMB effect on molecular-level populations and radiative transfer can result in errors of a few percent when estimating intrinsic line fluxes. For instance, using line fluxes, such as CO (1-0) or CO (2-1), to trace molecular gas mass can lead to inaccuracies in interpreting molecular gas properties, including total mass, density, and temperature.
 However,  this effect, which  is limited to millimeter/submillimeter wavelengths, cannot be addressed easily \cite{Zhang2016}.
 The impact of CMB temperature on high-z observations has been treated numerically in the work of \cite{Tunnard2017} in order to explain the observed deviations, at z$\gtrapprox 1.5$, from the Gao-Salomon relation, which strongly correlates the FIR and HCN(1-0) luminosities over more that 10 orders of magnitude in the local universe. \cite{Tunnard2017} conclude that the CMB is unlikely to explain the deviations reported in the literature, under reasonable conditions. However, the strength of the CMB effect is extremely sensitive to the kinetic temperature, density, and optical depth of the gas. 
 The CMB attenuation of HCN line intensities has also been explored by \cite{Rybak2022} as a possible reason for the scarcity of HCN  detections in  DSFGs at cosmic noon (see Sec. \ref{sec:dense}): an attenuation of $\sim 10- 30 \%$ was found for a DSFGs sample at z$\sim 3$, bringing to the conclusion that, up to z$\sim 3$, the effect of CMB on the HCN detectability is almost negligible. \\
Another aspect that should be taken in consideration is the potential impact of  enhanced dust temperature on the grain surface chemistry and on the desorption rates of species formed on grain ice mantles \cite{Fraser2001}. In the Milky Way, complex organic molecules have been detected in the gas phase of cold, prestellar  cores \cite{Bacmann2012}, suggesting that desorption mechanism can be effective also at low temperatures \cite{Potapov2021}.  To date, this effect on high-z galaxies has not been investigated yet.

\section{Interstellar radiation field} \label{sec:FUV}
Photodissociation regions (PDRs; also known as photon-dominated
regions) are responsible for most of the IR radiation in galaxies. They consist of predominantly neutral gas and dust illuminated by
far-ultraviolet (FUV) photons (6 < h$\nu$< 13.6 eV) of the interstellar radiation field. In PDR, heating and chemistry are dominated by the FUV photons \cite{Bisbas2012}, which are instead attenuated in the denser, more oscured inner parts of a molecular cloud, where the chemistry is initiated by cosmic rays, \cite{ODonoghue2022}, \cite{Wolfire2022}. 
It is frequently assumed that the higher the SFR, the higher will be the unattenuated intensity of the FUV radiation field, $\chi$ \cite{Bialy2020} (and, as  discussed in Sec. \ref{sec:cosmic_rays}, the CR ionization rate is expected to increase in a similar fashion).  However, attenuation is strongly related to the dust content per parcel of gas. 
Assuming a constant gas-to-dust ratio of 100,  and dust grains column density  scaling linearly with metallicity, \cite{Bisbas2023} shows that a visual attenuation A$_{\rm V}$ of a few tens can already  be attained at densities n$_{\rm H}= 10^5 $ cm$^{-3}$.  Thus, the dust component efficiently  shields the dense gas from the ionizing FUV field 
which, in high-z DSFGs , can be as high as $\sim 10^5 $ $\chi_0 $  \cite{Canameras2018}, \cite{Rybak2019} , being $\chi_0 $  the Draine FUV field  \footnotemark{} \cite{Draine1978}
The analysis of the FIR stacked spectrum  of high-z DSFGs presented in \cite{Wilson2017} is consistent with a value of the interstellar radiation field in PDR regions which is at least a factor of 10$^3-10^5$ larger than the Milky Way. 
Despite the higher rates of star formation and of FUV radiation with respect to quiescent galaxies, high-z DSFGs appear to be well shielded against  the FUV field just thanks to  the large dust-to-stellar mass ratio, which, for this galaxy population, lies above local spirals and ULIRGs \cite{Santini2010}, \cite{Donevski2020}. The consequences of the dust shielding on the molecular enrichment will be discussed  in Sec. \ref{sec:dust}.
\footnotetext{G$_0$ is the FUV radiation field in Habing units; G$_0=1$ corresponds to a flux of $1.6 \times 10^{-3}$ erg cm$^{-2}$ s$^{-1}$. As an example, the median value in the Milky Way is G$_0$=1.7, corresponding to a flux of $ 2.72 \times 10^{-3}$ erg cm$^{-2}$ s$^{-1}  \equiv \chi_0$.}  
\section{The far-reaching effect  of Cosmic Rays on star formation}\label{sec:cosmic_rays}
In general, the interstellar radiation field impinging on an a cloud is dominated by the UV radiation emitted by the more massive stars; as the most energetic photons are retained within the HII regions surrounding the brightest stars, the energy of the ionizing photons permeating the neutral ISM is lower than the Hydrogen ionization treshold of 13.598 eV and, importantly, lower than the ionization potential of many abundant species such as O, He, N and molecular Hydrogen. In general, these species are not photoionized in the diffuse clouds. The FUV radiation  field becomes strongly attenuated by dust as it propagates into a cloud. For this reason, only few species (such as Carbon and Sulfur) in the outer shells of a cloud and in the diffuse gas will be affected by direct photoionization, thus undergoing a quite limited chemistry. In these Photon-Dominated Regions (PDRs), the dominant source of free electrons is C$^+$ which, through associations with H$_2$ and dissociative recombinations, can form neutral C, CH, CH$_2$, CH$_4$. Importantly, the fast atom-radical reaction between O and CH can be a significant route to the important tracer molecule CO.  At depths of about two magnitudes in clouds in the average interstellar radiation field,  H$_2$ becomes abundant and CO becomes a significant reservoir of the available carbon. 

In the densest phase of the Interstellar medium, where external starlight is excluded, chemistry is instead initiated by the Cosmic rays (CR), which can penetrate large column densities of gas and dust and reach the molecular cloud core, maintaining, in the mostly neutral medium,  a small fraction of gas ionization which is sufficient to drive an extremely rich chemistry \cite{Indriolo2013} and to couple the dense gas to the local magnetic field. The  ionization fraction observed in the Galactic dense MCs ($\sim 10^{-7}$) can only be explained by CR.  \\
The Fermi Gamma-ray Space Telescope revealed that the diffuse $\gamma$-ray background is dominated by star forming galaxies \cite{Roth2021}. The $\gamma$-ray emission of a star-forming galaxy depends on its SFR, which determines supernova rate and thus the rate at which CRs are injected in the ISM, where CR collide with nuclei producing $\gamma$ photons.  This suggests that high SFR of high-z DSFGs, both  Main Sequence and Starbursts, harbor enhanced CR fluxes with respect to quiescent galaxies. Understanding the paramount role of cosmic rays (CR) in regulating the star formation efficiency of DSFGs is straightforward. An accurate characterization of their ionization rate can explain the mechanism fueling the extreme starbursts observed at the cosmic noon.
 Detailed reviews on the impact of CR in MCs  can be found in \cite{Gabici2022} and \cite{Padovani2020}. 
Since the cross section of the main molecular clouds processes (dissociation, ionization and excitation of H$_2$) peaks at relatively low energies, star formation processes are mainly affected by  low-energy CRs (E<1 TeV) \cite{Viti2013}, \cite{Gabici2022}, \cite{Padovani2009}, \cite{Padovani2018}, \cite{Padovani2020}, \cite{Padovani2023proceedings}.    \\
Below, we summarize the current status of Galactic and extragalactic observations and outline the primary impact of CR on astrochemistry and molecular gas. 

\subsection{CR Observations at different redshifts}
Observations through different  lines-of-sight in the Milky Way allowed to estimate the Galactic ionization rate of Hydrogen molecules by CR,  $\zeta _{H_{2}}$ as a function of the H$_2$ amount, covering H$_{2}$ column densities ranging from values typical of  diffuse clouds N$_{H_2} < 10^{21}$ cm$^{-2}$, up to the densest MCs N$_{H_2} \sim 10^{24}$ cm$^{-2}$. These estimates assumed a Galactic background of CR  propagating in the ISM, with an average value of $\zeta_{H_{2}} \sim 10^{-17} - 10^{-16} $ s$^{-1}$ (\cite{Padovani2009}, \cite{Indriolo2015}).   Galactic observations  have been  complemented with extragalactic data of the nearest starburst NGC 253 \cite{Holdship2022}, \cite{Harada2021}, ARP 220 \cite{Gonzalez2013}, Mrk 231 and, remarkably,  with two strongly lensed dusty galaxies at high z in the range $\sim 3.5-3.9$, made possible recently thanks to the first detection of H$_3$O$^{+}$ at such redshifts  \cite{Yang2023}. 
In most Galactic sources, the CR flux, as $\zeta _{H_{2}}$, follows the expected decrease with increasing column densities, as CR dissipatively collide with $H_2$  molecules. However, a large number of outliers are found,  that cannot be explained by the average Galactic CR flux: namely, in the Galactic Center \cite{Rivilla2022}, in a protostellar cluster \cite{Ceccarelli2014}, and in the vicinity of a Supernova remnant \cite{Ceccarelli2011},\cite{Vaupre2014}, where the estimated $\zeta _{H_{2}}$ overcomes the Galactic average by up to five orders of magnitude. These findings firstly opened the perspective for a local origin of low-energy CR acceleration. They suggest that the observed  spread in Galactic CR flux compared to the average Galactic value mirrors the  local  star formation state, and that the latter  is the primary source of low energy CR. 

However, the exact mechanisms responsible for their production remain debated, with proposals ranging from diffusive acceleration at supernova remnant shocks  \cite{Bell2013},\cite{Hong2023}, to protostellar shocks and jets \cite{Padovani2015}, \cite{Padovani2015}, or other mechanisms. \\
Noticeably, also the five observed Giant molecular clouds in the local starburst NGC 253 reveals ionizing rates  of about $1-80 \times 10^{-14}$ s $^{-1}$. Higher CR ionizing rates are, finally, also estimated for ARP 220, Mrk 231 and for the two high-redshift sources observed by \cite{Yang2023}, confirming that, up to high redshifts, star formation is directly traced by cosmic rays.  Going even further, the enhanced CR flux in SF regions  may be the leading actor in shaping the initial mass function towards top-heavy trend, because of the increased CR heating, at the expenses of an increasing destruction of molecular Hydrogen and of  a rich interstellar chemistry. The burst event itself may indeed prove to be a  self-sustaining process, underscoring the need for a quenching mechanism on the scale of the star-forming galaxy .

\subsection{Molecular signatures of CR dominated ISM } 
Since different chemical drivers (either UV-X photons or CRs) lead to different characteristic molecules through different chemical reaction networks,  one key aspect to consider when modeling integrated signals from unresolved  high-z DSFGs is that of  the unknown  filling factors of PDRs with respect to  CR-dominated regions. Viceversa, the observation of emission lines from  tracers of a specific chemistry will allow to infer the main chemical spark of the  ISM chemistry.  \\ 
Focusing on dense clouds, where the bulk of star formation occurs, we assume that the chemistry is started by CR ionization:  $\zeta_{H_{2}}$  is the parameter generally used to quantify the CR intensity while implementing astrochemical network. As shown in \cite{Viti2013},\cite{ODonoghue2022},  it strongly affects the physics and  chemistry of the ISM in many aspects, some of which is briefly summarized below:
\begin{itemize}
 \item {\bf Effects on the gas temperature}\\
 Cosmic rays produce ions and excited molecules, which can significantly heat the gas and produce temperature gradients in prestellar cores ( see e.g., \cite{Padovani2009}, \cite{Glassgold1973}, \cite{Dalgarno1999}, \cite{Glassgold2012} and the reviews \cite{Galli2015},\cite{Gudel2015}). In a molecular environment, the available energy goes into ionization of H$_2$, 
vibrational and rotational H$_2$ excitation, and in kinetic energy of the outgoing electron, available for secondary ionization. About 50$\%$ of the CR energy is lost in gas heating \cite{Glassgold2012}, \cite{Jonkheid2004}. 
Embedded protocluster can also accelerate CR from protostellar surfaces via accretion shock, producing CR ionization rates (and gas heating) higher than the average value of the host galaxy \cite{Gaches2018}. Due to the importance of the molecular gas temperature in the reaction rates of chemical reaction, CR heating cannot be neglected when modeling the physical environment for a chemical network.
\cite{Galli2015}, \cite{Glassgold2012}
\item {\bf H$_2$ dissociation and  H abundance in molecular clouds} \\
As discussed in \cite{Padovani2018} , secondary electrons from primary CR ionization contribute to H$_2$ dissociation, increasing the fractional abundance of atomic H and resulting in the only source of atomic H in dense clouds (N$_{H_2}>10^{21}$ cm$^{-2}$). This can severely alter the HCO abundance, the composition of grain mantles and the formation of complex organic molecules \cite{Tielens1982}. \\
\item    {\bf Enhancement of C/CO in dense clouds} \\
\cite{Prasad1983} showed that the CR penetration in dense molecular clouds can induce electronic excitation of the absorbing gas, particularly  H$_2$, resulting in the emission of a chemically significant UV flux. The latter can photodissociate the CO reservoir, adding to the CO destruction process by He$^+$ and recovering atomic neutral C, thus enhancing the abundance ratio C/CO in dense gas \cite{Bisbas2017}, \cite{Gaches2019}. \\
\item   {\bf Effects on dust grains }   \\
CR-induced UV photons can significantly alter the net  electric charge distribution in submicron grains \cite{Ivlev2015} , and regulate the photodesorption process on dust \cite{Oberg2009a},\cite{Oberg2009b}. 
During the lifetimes of about $4-6 \times 10^8$ yr of interstellar ices on  dust mantles  within dense molecular clouds, the long exposure to ionizing radiation (CR or CR-induced UV photons) can modify the pristine ices, favoring the formation of complex organic molecules  \cite{Bennett2007a}, \cite{Bennett2007b}, \cite{Oberg2009c}, \cite{Oberg2010} .  \\
\item  {\bf Ionization of H$_2$ } \\
The most remarkable effect is the ionization of molecular Hydrogen, which, through the formation of the trihydrogen cation $H_{3}^{+}$, initiates a chain of ion-neutral reactions that produce a large variety of chemical species. This pivotal ion is destructed in diffuse clouds by dissociative recombination , and in dense clouds by a proton-hop reaction with the CO molecule. The more direct estimate of $\zeta _{H_{2}}$ can be obtained my measuring the  $H_{3}^{+}$ column density (e.g. \cite{Oka2006}) and either the CO abundance (in dense clouds) or the ionization fraction and the H$_2$ fraction (for diffuse clouds), together with an estimate of the depth L of the cloud along the line of sight. \\
\item  {\bf CR effects on Oxygen chemistry} \\
In the diffuse clouds and at the edges of dark clouds, where there is still a non negligible fraction of atomic H, CR can ionize H, starting the Oxygen chemistry.  In dense clouds, the Oxygen chemistry follows the H$_{3}^{+}$ route, where the Oxygen network is triggered by the CR ionization of H$_2$. In both cases, through a number of charge transfer and abstraction reactions culminating with the dissociative recombination of H$_3$O$^+$ into water and Hydroxyl radical OH: the latter can then be used as a tracer of CR ionization rate, whenever CR dominates over UV ionization.  Intermediate ions of this reaction chain are the Hydride cations  OH$^+$, H$_3$O$^+$, H$_2$O$^+$: their use as CR tracers came into the game following the detection by the Herschel Space Observatory \cite{Gerin2010}, \cite{Neufeld2010},\cite{Indriolo2015}. \\
\item {\bf Hydrogen Deuteride HD} \\
In the CR dominated clouds,  CR determine H$^+$ formation which defines the D$^+$abundance through charge exchange reactions. HD is then produced by the fast ion-molecule reaction H$_2$+D$^+$ $\rightarrow$ HD+H$^+$   \cite{Black1973}, \cite{Hartquist1978}. HD is the main Deuterium reservoir in molecular clouds. The HD abundance can then be used to infer the CR ionization rate of atomic Hydrogen, which is slightly different than that for H$_2$ \cite{Glassgold1974}. \\
\item {\bf Deuterium fractionation}  \\
Deuterium fractionation starts with the formation of the protonated molecular Hydrogen H$_{3}^{+}$ and its isotopic exchange with the HD molecule which leads to H$_2$D$^+$. In low temperature (< 100 K) dense clouds, the endothermicity of the reverse reaction  unbalances the number density ratio $n$(H$_2$D$^+$)/$n$(H$_{3}^{+}$) towards much larger values  with respect to the cosmic elemental ratio $n$(D)/ $n$(H) $\sim 1.6 \time 10^{-5}$.   H$_2$D$^+$ is mainly destroyed by reactions with proton hop with CO, producing DCO$^+$ and by dissociative recombination.  Similarly, H$_{3}^{+}$ protonates CO to form HCO$^+$.  It is possible to relate the deuterium fractionation of HCO$^+$  , i.e. the ratio between the number densities $n$( DCO$^+$)/$n$( HCO$^+$) , to $\zeta _{H_{2}}$, making these two species important tracers of the CR effect on dense clouds \cite{Guelin1977}, \cite{Caselli2002}, \cite{Ceccarelli2014d} \\

\end{itemize}

\section{Dust and metallicity environment for DSFGs molecular clouds}\label{sec:dust}
A key parameter of any astrochemical modeling of molecular enrichment, is the metallicity of the atomic environment out of which molecular clouds originate. High gas-phase metallicities promote the chemical  production of oxygen-bearing and carbon-bearing molecular species, and, importantly, set the  physical environment for dust production. Metals are added, removed and redistributed in the ISM  by star formation, supernovae and winds, and are partially depleted off the gas phase through  the condensation into dust grains.
 Estimates of metallicity (Z) at high z are always extremely complicated and strongly affected by systematics due to calibration methods, which can vary the metallicity estimates even by 0.7 dex \cite{Kewley2008}. What is nowadays accepted by the scientific community is that, from a statistical point of view, metallicity increases with the (total)  stellar mass and decreases for increasing redshifts.  The rapidity of this decrease, and its causes, are not yet completely understood. First observations pointed to  a rapid decrease of Z at increasing redshifts, for a fixed stellar mass. More recent works seem to suggest a slower decrease \cite{Sanders2021} and, importantly, that this decrease is related to the SFR rather than to the redshift.  In other words, for a fixed stellar mass, galaxies with progressively higher SFR  are selected at increasing z , and the SFR is observed to be anticorrelated with metallicity \cite{Mannucci2010}, \cite{Curti2020}: the Z evolution with redshifts can thus be described in terms of SFR. \\
We outline that the above picture gives a statistical view: the {\it average} cosmic metallicity decreases at higher redshifts because the average is dominated by many “small” (low total stellar mass) and metal-poor galaxies. \\
The situation is different when focusing on single objects (as DSGFs), since, ultimately, the metallicity is regulated by the combination of inflows of metal-poor gas, outflows of enriched material, and star formation.  Almost all models predict an initial rapid increase in metallicity, up to the achievement of a saturation level, a sort of equilibrium value of Z where inflows, outflows and star formation are balanced \cite{Lilly2013}, \cite{Pantoni2019}, \cite{Lapi2020}. This equilibrium value depends on the metallicity of the inflow, on the mass-loading factor (which measures the efficiency of outflows in removing gas from the galaxy relative to the formation of stars), and on the SFR (the relative importance of these factors in determining Z has  been the subject of numerical simulations in \cite{Bassini2024}). The point is that, in general, at any redshift  there are, as a matter of fact, some galaxies which are more dusty and more metal rich than the Milky Way, and this is not contrast with the cosmic trend, mentioned above, of an average metallicity decreasing at increasing z.  As an explanatory example, we mention the observations of a sample of 30 distant quasars (z$\sim 6$) by \cite{Juarez2009},  in which the inferred metallicity of  the host galaxies is as high as several times solar. \\
The value of gas metallicity in high-z DSFGs is still matter of debate \cite{Tran2014}, \cite{Liu2008}, \cite{Donevski2020}. This is because, in heavily dust-enshrouded galaxies, metallicity diagnostics relying on rest-frame optical emission lines are not usable. However, observations  of the FIR fine-structure emission lines [NII]205 $\mu$m, [CII]158 $\mu$m \cite{Nagao2012}, [OI]145 $\mu$m , [OIII]158 $\mu$m \cite{DeBreuck2019}, and  of H$_{\alpha}$ and [NII] 658.4 $\mu$m \cite{Birkin2023}, suggest that SMGs at $z \sim 4$ are already chemically enriched nearly to solar metallicities, as a result of a rapid metal enrichment in the early phase of star formation. Recently,  JWST observations delivered the first spatially resolved maps of gas-phase metallicity for two gravitationally lensed dust-obscured star-forming galaxies at z$\sim 4$ \cite{Birkin2023}, showing that dust surface density and gas surface density  have spatial variations positively correlated with metallicity. This indicates that regions containing more gas and dust are also more metal rich. The spatially averaged metallicity in the sample of \cite{Birkin2023} is conservatively estimated to have a value of Z$\sim 0.7$ Z$_{\odot}$. Of course, the average galactic metallicity is expected to follow the SFH and evolve in a timescale set by galactic dynamics \cite{Torrey2018} , while the evolution of gas metallicity within the single molecular cloud evolves in the shorter timescales  of the cloud lifecycle \cite{Chevance2020}. A star forming galaxy  is expected to host molecular clouds at different stages of their evolution, from starless, dark clouds to protostellar cores. These clouds may originate from atomic gas with differing metallicities. The statistical distribution of clouds evolutionary stages, and of the environmental metallicity, should be taken into account when modelling the integrated signal from high-z DSFG.   \\
 Besides the gas metallicity, an important parameter needed to model the molecular enrichment in a cloud  is the value of the M$_{dust}$/M$_{gas}$ ratio,  $\delta_{\rm DGR}$, as it determines the opacity in a parcel of gas, thus regulating the chemical driver of molecular formation, as well as the amount of grains available for surface chemistry.   We know that high-z DSFGs are dust rich, with  dust-to-stellar mass ratios (M$_{dust}$/M$_{\ast}$)  higher than in  spiral galaxies (by a factor of $\sim $ 30) or ULIRGs,  and that they are gas rich, with gas fractions (M$_{gas}$/M$_{\ast}$) approaching 50$\%$ \cite{Tacconi2008}:  indeed,  the high M$_{dust}$/M$_{\ast}$ ratio  plausibly mirrors their high gas content \cite{Santini2010}, \cite{Donevski2020}.  
The findings outlined in  \cite{Birkin2023} indicate that  within each  resolved region of the observed DSFGs, the dust-to-gas ratio tends  to raise with higher metallicity. This trend mirrors findings observed in local galaxies. Additionally, when $spatially \ averaged$, this ratio is found to be moderately lower than the canonical  value typically adopted for local ULIRGs ($\delta_{\rm DGR} \sim 1/100$, \cite{Leroy2011}) , a result also discussed in \cite{Donevski2020}. What is relevant for our purposes is to note that, taken individually, the resolved ($\sim 0.7 $ Kpc ) 
regions of \cite{Birkin2023} span about one order of magnitude in  $\delta_{\rm DGR}$, confirming that molecular clouds, with sizes spanning the range of 10-100 pc, are caught in different environments/evolutionary stages, where timescales for evolution refers to the molecular cloud lifecycle (10-30 Myr, \cite{Chevance2020}) : no single astrochemical model can fit the whole galactic signal, making necessary to use a statistical approach in the astrochemical modeling of high-z galaxies molecular abundances. 
The dominant provider of dust remains unclear: whether it is AGB stars,  SNe, star-forming regions, or the processing coagulation and growth of dust facilitated in metal-rich gas,  is yet to be determined \cite{Matsuura2009}, \cite{Gall2018}, \cite{Morgan2003}. 

We know that dust is a fundamental player in the physical and chemical processes occurring in the ISM (see, e.g., \cite{William2002}).  In the ISM, dust grains show up as bare grains of refractory materials or covered by an ice mantle, depending on the environment. In diffuse clouds (50-100K), the effect of the UV radiation favors the presence of bare grains, whereas ice mantles are typical in colder, shielded clouds.  It is widely believed that dust grains divide into two main classes, depending on their chemical composition: one carbon based, and one, dubbed “astronomical silicates” , dominated by O, Fe, Si and Mg. Usually, the silicate class is represented with one elemental partition of olivine, namely MgFeSiO$_4$.  
The most studied effects of dust in affecting galaxy evolution are the reprocessing of stellar radiation and the depletion of ISM metals (e.g. \cite{Draine2003}, \cite{Jenkins2009}). However, detailed dust properties, such as composition and size, can only be constrained by measured extinction curves along different sightlines, which are available, to date, only for the Milky Way and the Magellanic Clouds (e.g. \cite{Nozawa2013}). Furthermore, these basic properties depend on the environment (i.e., they are not homogeneous throughout a galaxy’s volume) and evolve with time. On the other hand, the composition and, even more, the size distribution of dust grains strongly affect the molecular formation rates through surface astrochemistry: for a fixed total  mass of dust grains, the smaller grains  provides a larger surface area where molecules can form  (in particular, H$_2$),  rapidly diffuse and react, acting as a third body to dissipate the energy released in exothermic bond formations, and catalyzing reactions needing  to  lower the activation barrier (e.g. \cite{Weingartner1999},\cite{Potapov2021}) .  
Only in the last decade, a complete treatment of grain size distribution for carbonaceous and silicate dust began to be investigated through semi analytical or numerical simulations \cite{Asano2013}, marking the beginning of the “dust cycle” models. \\ The general trend arising from the first  works is that dust {\it seeds}, produced via stellar channels (outlined later in this Section), evolve from an initially distribution dominated by large grains ($\gtrsim 0.1 \mu$m) to a broad size distribution, numerically dominated by the small radius  end. The size evolution can be followed reasonably well just considering two representative sizes, referred to as “large grains” and “small grains” , whose limiting radius is set at $\simeq 0.03  \mu$m \cite{Hirashita2015}. This simplification  allowed the development of more sophisticated numerical simulations of dust evolution (e.g. \cite{Granato2021}, \cite{Parente2022} and references therein). 
The  dust  cosmic lifecycle mainly starts in the ejecta of evolved stars of mass $\lesssim$ 8 M$_{\odot}$ ( the asymptotic giant branch , AGB,  stars), and in their later evolutionary stages, in planetary nebulae and in SNII (possibly also in SNIa explosions) . 
In the hot circumstellar atmospheres of the AGB stars, the primary dust formation, forming  grains with sizes from a few to tens of nanometers,  occurs in gas phase: the atmosphere  is rich in Carbon, Oxygen and Silicon, which form in the stellar core and are ultimately dredged up to the stellar surface thanks to mutiple convections of the stellar hot plasma. In oxygen-rich AGB stars atmospheres, where n(O)/n(C) > 1 , Carbon is predominantly trapped in CO: the formation of carbonates requires that three oxygen atoms are available for each C atom in the carbonate, but the extremely high bond energy of the CO molecule implies that this  cannot happen in a carbon rich element mixture if  the oxygen is completely consumed by CO formation,  so that  oxygen atoms react with silicon and any other metals forming  amorphous and crystalline oxide and silicate grains \cite{Ferrarotti2005}, \cite{Rimola2021}.  Instead, Carbon-rich stars, where the ejecta have n(O)/n(C) < 1,  are assumed to condense only  graphite or amorphous carbon grains. Thus, primary grains consist of either carbonaceous materials \cite{Ehrenfreund2000}, \cite{Tielens2008} or silicates (refractory materials, \cite{Henning2010}), depending on the C/O ratio in the gas phase where dust grains nucleate and form: the production of carbonates or silicates is {\it mutually exclusive} in AGB winds. 
In a single stellar population, already after $\sim$ 5 Myr stars with $\sim 8 $ M$_{\odot}$<M< $\sim$ 50 M$_{\odot}$ undergo core collapse in a Type II Supernova.  In Type II SN, dust grains can nucleate  and condense already in the heavy-element rich mantle: in a onion-like model, there is no molecular intermixing between C and O layers, and Sne may produce both carbonates and silicates grains \cite{Kozasa1989}.  Conversely from AGB stars, in the outflows produced by SNae II and Ia explosions, carbon and silicate dust can condense at the same time \cite{Dwek1998}.
The classical picture is that the main contributors to carbonated dust are low mass AGB stars \cite{Marigo2001}, whereas SNe II are the main contributors of silicates in the ISM \cite{Woosley1995}. It is, however, important to outline that the subsequent thermal pulsations in AGB stars (due to the explosive ignition of the Helium shell)  causes multiple phases of dredging up, able to turn an O-rich giant to a C-rich one, and that, in general, the overall yield of dust in SNae and AGB stars is still highly unknown. For example, detailed nucleation calculation by \cite{Ferrarotti2005} and \cite{Morgan2003} shown that the dust yields from AGB stars is largely overestimated \cite{Dwek2005}.
In any case, dust cycle starts with the ejecta of AGB  stars or SN,  the primary grains or  “dust seeds”, consisting of large grains  \cite{Bianchi2007}, \cite{Nozawa2007}.  Once they are injected in the ISM, primary grains undergo a series of physical processes able to change both their composition and their size:
Gas metal atoms can stick to the surface of the grains. Being a surface process, this accretion is particularly relevant for small grains, having larger total area per unit mass. $Accretion$ changes size and composition of small grains. 
Grain-grain collisions can produce $coagulation$ (when the collision occurs at low velocities, such as in the dense, cold  ISM) or  $shattering$ (for high-velocity impact) expected in the diffuse ISM: the former acts   a source of large grains and a sink of small grains, and vice versa for the shattering. 
$Sputtering$ is another surface process, thus relevant on small grains, due to the erosion of the grains by ions in the gas phase. This erosion can be due to collision with energetic ions and by SN shocks, with extreme effects bringing to total grain destruction, and dominates the hot plasma at T$\gtrsim 10^5$ K. 
Astration is the last step of the dust cycle, returning processed dust to the stellar component. \\
It is obvious that the size distribution and composition, at a given evolutionary stage of a galaxy, is not easily predictable and requires accurate simulations of the above-mentioned opposing effects.  The most recent simulations by \cite{Parente2022}  refer to objects with Dark Matter halo mass up to 3 times that of the Milky Way (which is $\sim 10^{12}$ M$_{\odot}$), showing that, at a representative “cosmic noon” redshift of z $\sim$ 3, or lookback time of $\sim $ 12 Gyr, the fraction of small to large grains is $\gtrsim 10^{-2}$, with a low dominance of silicates over carbonates. But high-z , massive DSFGs dark halo may be even 10 times larger than the values adopted in that simulation. 
So, the size distribution between large and small grains, and their compositions, in DSFGs, and, more generally, in high-z galaxies, is still an open issue. As for the high-z DSFGs, it may be that the large SFR, the enhanced SN rate and cosmic ionization rates, the possible high fraction of diffuse vs. dense gas (see Sec. \ref{sec:dense}) makes a big part of primary grains to evolve into small grains. This hypothesis needs, however, to be supported by more observations and/or detailed simulations on small scales. 
 Noticeably, the most recent observations by JWST are providing interesting challenges to the dust-cycle picture, while exploring the most extreme distances observed in the galactic framework. An absorption feature around a rest-frame wavelength of $\lambda=2,175 \AA$ was found in the spectrum of a galaxy at z=6.71 \cite{Witstok2023}, i.e. just 800 Myr after Big Bang. This feature, known as the ultraviolet (UV) bump, was first discovered along the lines of sight in the Milky Way and generally attributed to small carbonaceous dust grains, specifically PAHs (size 0.3-5 nm) or nano-sized graphitic grains. The puzzle resides in the fact that, at such redshifts, AGB stars haven’t formed yet, so that the only plausible source of dust seeds is SNe \cite{YangLi2023}, which in the local Universe, are more prone to produce silicates in larger amounts than carbonates: the latter are, however, a non-negligible SN ejecta. On the light of the processes summarized above, we may argue that the “down-sizing” fragmentation processes acting on the large seeds grains   ($\sim 0.1 \mu$m) to produce nano carbons or PAHs ($\sim nm$) must be much effective at those high redshifts, by a combination of shattering possibly followed by sputtering. This investigation goes well beyond the purpose of this paper, but make us reflect on possible scenarios justifying the efficient formation of PAHs  in high-redshift sources.

Whatever the main factory is, dust plays a paramount role in the ISM molecular enrichment,  both for its thermal effects on the gas,  and for the several chemical processes occurring on the  dust grains surfaces.  Dust regulates the balance of heating and cooling in the ISM, as the massive gas reservoir, and the efficient dust shielding from the FUV field, described in Sec. \ref{sec:FUV},  implies that a large amount of gas, in the form of diffuse clouds (atoms and simple diatomic molecules, with characteristic temperatures of 30-100 K and gas density n$_H \sim 10-5 \times 10^2$ cm $^-3$)  will undergo the necessary cooling for gravitational collapse, allowing the formation of dense, cold gas clouds with temperatures of $5-10$ K and n$_H > 10^4$ cm $^-3$ ,   pristine cradle of star forming regions. 
\subsection{Impact on molecular chemistry}
In the diffuse gas, dust grains are solid-state, sub-micron sized, bare silicates and carbonaceous refractary compounds, while prestellar grains (in dense cores) are coated with thick ice mantles, cradles of complex organic molecules. We  summarize below the main roles of dust on molecular chemistry. 
\begin{itemize}
\item{ \bf Traffic warner of  chemical drivers} 
The dust-to-gas ratio and the chemical composition and size of dust grains will result in the visual attenuation A$_{\rm V}$  against the FUV interstellar field, thus determining the “thickness” of the PDR regions, layers of neutral gas separating photon dominated chemistry from CR dominated chemistry, as discussed in Sec. \ref{sec:FUV} (see also \cite{Viti2013},\cite{Byrne2019},\cite{Yang2023}). For A$_{\rm V} >1-2$, Carbon is mainly in neutral form, and increasingly incorporated into CO molecules for increasing depths in the cloud. While the chemistry in PDR regions is limited to the formation of few simple molecular species, the transition to dense cores comes together with a very rich molecular scenario  \cite{ODonoghue2022}. For this reason, for a given interstellar FUV field, the dust-to-gas ratio and the grain composition are of paramount importance in regulating the overall filling factor of the PDR regions with respect to dense molecular cores. 
\item{\bf Ruler of atomic-to-molecular transition} 
In dense clouds, the atomic-to-molecular transition is regulated by the balance between the formation of  H$_2$ on grains surfaces and on the H$_2$ destruction by the FUV external radiation field. Increasing the optical depth in the cloud, photodissociation is reduced by dust and by H$_2$ self-shielding (through absorption in the Lyman-Werner bands). The formation rate is proportional to the gas density, thus the primary controllers of the transition from atomic to molecular gas are the dust-to-gas ratio or the metallicity of the gas (which determine the dust absorption opacity), the gas density and the intensity of the FUV field \cite{Tacconi2020}. 
\item{\bf Factory of H$_2$, H$_2$O, O$_2$, and complex organic molecules}
Being H$_2$ the most abundant molecule in the ISM, it  gives an important contribution  to the cooling of collapsing gas, necessary for star formation. The primary route for the H$_2$ formation is grain-surface chemistry \cite{vanDeHulst1946}, \cite{Hollenbach1970}, \cite{Cazaux2004}, \cite{Vidali2013a}, \cite{Vidali2013b},\cite{Vidali2013c}.  \\ The use of grain surface chemistry is also required to explain the observed abundances of water in molecular clouds \cite{DHendecourt1985}, \cite{Hasegawa1992} because of the inefficiency of the gas phase routes for its formation.  In the cold phase of the collapsing gas, dust grains are covered by thick  icy mantles. The prestellar grain surface chemistry is dominated by hydrogenation processes: simple hydrogenated molecules, like H$_2$O \cite{Cazaux2010}, \cite{VanDishoeck2013} and CH$_3$OH \cite{Watanabe2008},\cite{Hama2013},  form in this phase by hydrogenation of O, O$_2$, O$_3$ and CO. Molecules formed on icy grain mantles during the prestellar phase remain frozen in the grain mantle until the densest, central core of the collapsing cloud starts to heat up when forming a protostar, a central hot core with temperatures of about 100-300  K.  During this protostellar switch-on phase, grain-surface chemistry is thought to be responsible, together with gas-phase processes,  for the formation of many complex organic compounds, the so-called “interstellar Complex Organic Molecules” (iCOMS), carbon-bearing molecules with at least six atoms (see \cite{Ceccarelli2023} for a review).  Many common organic compounds are thought to form in this phase, such as methyl formate  HCOOCH$_3$, formic acid HCOOH, and dimethyl ether CH$_3$OCH$_3$.  The relative importance of grain surface and gas processes may increase with the duration of the warm-up phase from prestellar to protostellar core \cite{Garrod2006}, although more recent observations found that  iCOMS are numerous and relatively abundant (fractional abundance as large as $\sim 10^{-10}$) already in the cold phase, before the switch-on of the protostar \cite{Bacmann2012}, \cite{Cernicharo2012}, \cite{Walsh2014}. Finally, the warm-up phase generates ices sublimation, which injects the icy mantle molecules in the gas phase, where they could be detected though their rotational lines. The evolution continues with dissipation of the protostellar envelope, converted in protoplanetary disk.  \\
Although we can expect high-z DSFGs to be a huge resource of iCOMS, unfortunately the low scales  of hot corinos and hot cores are very compact (size less than 0.1 pc), and the strongest transitions of these large  molecules are located between $\sim $ 30–50 GHz rest frame frequency, which makes their direct detection in high-z DSFGs out of reach of the current astronomical facilities (but see, e.g. \cite{Codella2015} for the observational perspectives of the Square Kilometer Array SKA). However, it has been estimated by \cite{Lintott2005} if the number of hot cores are a fator of $\sim 1000$ larger than in the Milky Way, specific signatures of hot core chemistry may be detectable even at high z.  \\
Despite the chemical networks for surface reactions are still being developed by the astrochemical community,  it is now  widely accepted that  chemistry on dust can have an high impact on the molecular composition of the ISM gas. In particular, the high ionization rates and FUV fluxes of DSFGs are expected to play a role in the non-themal desorption from grain mantles. This is why, when dealing with high-z DSFGs and with their large dust content and extreme environmental conditions, surface chemistry should always be considered together with the gas routes for molecular formation. 
\end{itemize}


 \section{The molecular gas reservoir:  H$_2$ as traced by CO}
\label{sec:h2co}


The fundamental properties of galaxies traditionally adopted to trace their star formation cycle are the star formation rate (SFR), the stellar mass M$_{\ast}$ and the molecular gas mass M$_{H_{2}}$. They are connected by three scaling relations.  The Kennicutt-Schmidt (KS) relation \cite{Schmidt1959}, \cite{Kennicutt1998},  assumes that star formation is fueled by molecular gas \cite{Wong2002}, and illustrates how an increase in molecular gas corresponds to an increase in the star formation rate. The Molecular Gas Main Sequence (MGMS) \cite{Lin2019} shows that an increase in molecular gas is correlated to an increase in the stellar mass: a plausible interpretation is that  higher stellar masses trace higher gravitational potentials, which retain higher gas amounts, fostering  the conversion from atomic to molecular Hydrogen through enhanced pressure and densities.   Finally, the Star Forming Main Sequence (SFMS)  \cite{Brinchmann2004}, \cite{Whitaker2012},\cite{Renzini2015},\cite{Sandles2022}, illustrates the observed correlation between the stellar mass and the star formation rate. 
DSFGs form a heterogeneous class of galaxies: for the majority of known DSFGs,  SFR and M$_{\ast}$  follow an  approximately linear relation along the SFMS. Positive outliers,  having boosted specific SFR  (sSFR $\equiv$ SFR/ M$_{\ast}$) form the so-called Starbursts (SBs).  The interpretation of these characteristics within the main-sequence paradigm provides key constraints on the history of  DSFGs  stellar mass assembly  \cite{Daddi2010}, \cite{Rodighiero2011},  \cite{Sargent2014}, \cite{Scoville2017},\cite{Pearson2018}, \cite{Silverman2018}.
Molecular gas in high-z DSFGs is warmer than in less active objects, and has column densities higher than quiescent galaxies. A  negative correlation is observed between molecular gas depletion time and excitation (e.g., \cite{Paglione1997}), and  a positive correlation is found between gas density and star formation rate (e.g., \cite{Gao2004}).  This shows that  density and temperature of the molecular gas are related to star formation activity: the average densities and temperatures of the molecular clouds in starburst nuclei are higher than those of more quiescent galaxies. 
There is still debate in literature on the nature of the main scaling relations, on whether they are intrinsic and fundamental, or  some of them are  just indirect by-products  of the others (\cite{Daddi2010}, \cite{Genzel2010},\cite{Baker2023}). But, whatever their nature is, we know that the characterization of the star formation process needs to start from a reliable estimate of the molecular gas mass and, ultimately, of the molecular Hydrogen content. 
Molecular hydrogen is the  most abundant molecule in the universe and plays a central role in the evolution of stellar systems and galaxies.  It is found in all regions where the attenuation A$_V > 0.01-0.1$ mag, shielding the gas  against the ultra-violet (UV) photons responsible for its photo-dissociation (requiring an energy of about 12 eV).  However,  H$_2$ is a diatomic homonuclear molecule, and it has no permanent dipole moment and no dipolar rotational transitions. All ro-vibrational transitions within the electronic ground state are quadrupolar with low Einstein coefficient for spontaneous emission, emitting radiation very weakly \cite{Sternberg1989}, \cite{Habart2005} . More importantly, the two lowest para and ortho purely rotational quadrupole transitions are only excited at temperatures higher than few hundreds K: the cold molecular component of ISM is thus invisible in H$_2$ emission at the characteristic temperatures of typical Giant Molecular Clouds (GMCs) (10-20 K, \cite{Bergin2007}). The strong dipole-allowed H$_2$  transitions in the Lyman
($B^{1} {\Sigma_{u}}^{+}-X^{1}{\Sigma_{g}}^{+}(v',v'') $)
and Werner
($C^{1} {\Pi_u}  -X^{1}{\Sigma_{g}}^{+} (v',v'')$)
systems fall in the range of the vacuum ultraviolet (10-200 nm).  \\
Due to the difficulties in having a direct measure of the H$_2$ content, Carbon monoxide is routinely used as an observable tracer of H$_2$  \cite{Bolatto2013}, being the second most abundant molecule in the ISM, with the advantage of a weak permanent dipole ($\mu \sim 0.11 D$) and a ground rotational transition CO (1-0) of $\sim 5.53 $ K ($\lambda= $2.6 mm, rest frame frequency 115.22 GHz). 
The CO  ground transition has  critical density  for collisional excitation of n$_{crit}= 3.9 \times 10^2$ cm$^{-3}$), further  reduced by radiative trapping, due to its high optical depth.  The low excitation temperature and critical density  makes CO(1-0) a relatively strong, easily excited millimeter (mm) emission line, traditionally used to trace the bulk of the  molecular content. With respect to mid- and high-J CO transitions, it is the least affected by the excitation conditions of the gas, thus fully revealing the widely distributed reservoirs of less dense, sub-thermally excited gas (e.g. \cite{Papadopoulos2001}, \cite{Carilli2010},  \cite{Ivison2011} ).   In  high-z star-forming galaxies, observations of CO(1–0) require observing capabilities in the 20–50 GHz regime (wavelength 6-20 mm). 
The standard methodology to infer M$_{H_2}$ from CO assumes a simple relationship between the H$_2$ column density, N$_{H_2}$,  and  the integrated line intensity W(CO) of the $J=1 \rightarrow 0 $ purely rotational transition of the most common isotopologue $^{12}$C$^{16}$O (hereafter, CO). In nearby clouds throughout the Milky Way disk, the CO-to-H$_2$ conversion factor shows little variations and is given approximately by  \cite{Dickman1978}, \cite{Sanders1984},\cite{Dame2001}:
\begin{equation}
X_{CO, gal}= \dfrac {N_{H_2}} { W_{CO}} = 2 \times 10 ^{20} \ \rm{cm}^{-2} \ \rm{K}^{-1} \ \rm{km}^{-1} \ \rm{s}
\end{equation}
As a corollary,  integrating over the emitting area, M$_{\rm mol}= \alpha_{CO} {\rm L_{CO}}$, with $\alpha_{CO}= 4.3$ M$_{\odot}$  (K$^{-1}$ km$^{-1}$  s  pc$^{-2}$) .
Independent estimates of the molecular mass in local observations  usually come from virial mass measurements, or from measurements (or assumptions) about the dust-to-gas ratio (where the dust mass is estimated from the IR SED of the dust component, when available) , and $\gamma$-ray observations of the bremsstrahlung radiation emitted by the cosmic rays interaction with the dense molecular gas \cite{Strong1994}. 
Significant deviations from the typically observed relative constancy of the Galactic X factor
have been found in regions of high SFR density, where  X$_{\rm CO}$<  X$_{\rm CO, gal}$  by a factor $\sim $4-20 (e.g. \cite{Downes1998})  in nearby mergers and starbursts.  For long time,  the custom in literature  has been to assume a bimodality in the CO-to-H$_2$ conversion factor, with ${\alpha}_{CO} \approx 4 $ for normal star-forming system (similar to the typical value of local  Giant Molecular Clouds) and ${\alpha}_{CO} \approx 0.8 $ for boosted star formation systems.  When considering high-z systems, however, the combined measurement of dynamical mass and high-resolution CO(1-0) observations leads to a broad range of X factors, from lower to higher values compared to local ULIRGs (e.g. \cite{Tacconi2008}, \cite{Daddi2010}, \cite{Hodge2012}), with no evidence of a strict bimodality, nor of a single conversion factor applicable to high-z DSFGs.  \\
If one sticks to the forementioned  bimodal conversion factor, the Kennicutt-Schmidt (KS) relation between SFR density and  surface density of the cold molecular gas as traced by CO (1-0), separates in two distinct  regimes, where ULIRG and high-z SB lie above the more quiescent galaxies \cite{Casey2014}. This appears  to reflect the bimodality in the SFMS between MS and SB.  In turn, the SFMS  estimates the stellar mass in obscured star formation environments by assuming a given relation between stellar mass and gas mass, and, ultimately, assuming  a constant dust-to-gas ratio.   Viceversa, accurately modeling X$_{CO}$ as a smoothly varying function of the gas surface density, so accounting for the boosted W$_{CO}$ due higher velocity dispersions in heavily star-forming environments \cite{Feldmann2012a}, \cite{Feldmann2012b}, \cite{Glover2010},  the KS relation becomes unimodal. 
The conclusion is that, in  order to use CO(1-0)  as a proxy for H$_2$, some caution must be adopted, especially in the extreme ISM conditions of high-z DSFGs. The “X factor” is explicitly dependent on the H$_2$  column density, the peak CO intensity (brightness temperature) and range in velocities, and, implicitly, dependent on density and kinetic temperature. 
In the physical environment of high-z  DSFGs, how does  the chemistry   affect the relative abundances of H$_2$ and CO and the value of the conversion factor? \\
In the diffuse molecular medium, dominated by photochemistry, both CO and H$_2$ are easily dissociated by FUV photons, making the abundances of these molecules small in the low density (n$_H$= 100-500 cm$^{-3}$ ) and low extinction (A$_{V}  \lesssim 1$ ) gas \cite{Snow2006}. In low extinction regions of the clouds, photoelectric heating is dominated by photons with energies above 6 eV, while photons with energies above 11.2 eV and 11.5 eV are responsible of H$_2$ and CO dissociation, respectively. \\
However,  H$_2$ can exist at lower column densities than CO because, being more abundant, it is more easily self-shielded than CO \cite{VanDishoeck1989}, \cite{Draine1996}, \cite{Lee1996}. There is a range of column densities ($10^{20}-10^{21}$ cm$^{-2}$) where the gas is potentially undetectable with the CO line (“CO-dark clouds”), so that the use of the Galactic X$_{CO}$ factor would underestimate the total molecular mass if the filling factor of these ISM component in non negligible.  This range of column densities is dependent on the thickness of the transition region and, ultimately, on the intensity of the FUV field. \\
CO photodissociation is also more effective in low metallicity environments: in this case, also the H$_2$ abundance may be reduced, but its higher self-shielding makes H$_2$  less sensitive to changes in metallicity.  In low metallicity systems, CO traces the regions of a cloud that have extinction greater than $\sim 2$ but does not trace the surrounding diffuse envelope, which can  be CO-dark while containing a large fraction of molecular gas \cite{Glover2011} , \cite{Hu2021}.  For the typical dusty and heavily star-forming systems at hand, we can expect that the metallicity plays a negligible role in boosting the CO photodissociation, while we argue that the relative size of the CO emitting region in a molecular clouds shrinks because of the boosted value of the environmental FUV associated to the high SFR. \\
It is  reasonable  to assume that, for what concerns the temperature and chemistry of the gas,  the relevant FUV interstellar radiation field comes from massive, young stars, so that the FUV intensity scales linearly with the star formation rate. Similarly, if cosmic rays are associated to SN remnants, the ionization rate $\zeta_{CR}$ can be assumed to track the SFR.  Moving to environments which may be typical of DSFGs,  in which molecular clouds are embedded in  strong FUV and CR fluxes, more CO photodissociation will occur, decreasing the filling factor of the CO emission and, if CO is not distributed throughout all the cloud volume, decreasing the linewidth, leading to an overall increase of X$_{CO}$ with respect to the Galactic value. On the other hand, the higher photodissociation and the higher CR heating in low A$_V$ regions will increase the brightness temperature of the CO emission, decreasing X$_{CO}$. \\
In order to determine the total effect on the conversion factor in different environmental FUV fields and CR fluxes, \cite{Clark2015} presented a series of simulations which incorporates  a time-dependent astrochemistry network into a smoothed particle hydrodynamics code,  finding that for virialized clouds with masses of $10^5 $ M$_{\odot}$ and mean density of $100$ cm$^{-3}$, the X$_{CO}$ factor increases by one order of magnitude over a SFR increase of two orders of magnitude.  Remarkably, the inclusion of CO destruction by dissociative charge transfer with He$^+$, occurring for large CR ionization rates, leads to a dependence on SFR even in the portions of the cloud which are highly shielded and CO bright.  In general, for high column densities (mean extinction A$_V$ above $\sim 6$ )  X$_{CO}$ becomes almost independent  on the external FUV field \cite{Feldmann2012}  but retains a dependence on the CR ionization rate.  \\ 
These simulations did not account for turbulence and density effects, though, and may be altered if the  clouds in  DSFGs are systematically denser and/or more turbulent than less actively star forming galaxies: while high SFR tend to increase  X$_{CO}$  (reducing the CO abundance), higher densities and turbulent velocities may counterbalance this effect by increasing W$_{CO}$. At difference with the diffuse gas, the high  density core of the molecular clouds are less affected from the external FUV field, and density and temperature play a major role: this has to be taken into account for molecular gas in high-z DSFGs, which is  on average warmer than in less active objects, and has column densities higher than more quiescent galaxies. \\
These effects were analyzed by \cite{Shetty2011}. Observations of Galactic CO show that, due to the large  optical thickness of the CO (1-0) line, W$_{CO}$ saturates beyond a threshold column density \cite{Lombardi2006}, \cite{Pineda2008}. When CO line saturates, it no longer traces gas mass, and W $_{CO}$ scales linearly with  N$_{H_2}$ . In the Milky Way, this saturation occurs at N$_{H_2} \lessapprox 10^{21}$  cm$^{-2}$. In general,  the CO line becomes saturated in regions with the highest CO abundance, so the corresponding H$_2$ column density depends on $f_{CO}=n_{CO}/n_{H_2}$ ratio which, in turn, is related to the level of CO destruction by CR and, ultimately, on the SFR. Also, the high dust content of DSFGs may alterate this ratio by increasing the shielding of CO.   \\
Given all the physical parameters involved, and the still unclear indications from simulations,  it is clear that much caution is requested when extrapolating the local X$_{CO}$ to the extreme  environments of  high-z DSFGs, where our interpretation  of Kennicutt-Schmidt relation should be rethought.  
\subsection{Other CO rotational lines}
Although  CO(1-0) could provide, with some caveats, a robust estimate of the bulk of the extended, low-excitation molecular gas reservoir, it is an intrinsically faint line, whose detection in  high-z DSFGs involve large observing time. 
The situation with high-z DSFG is particularly murky also  because the ground state CO transition is redshifted out of many typical instrumental bandpasses. For this reason, mid- and high-J CO transitions are often used: they are typically brighter than the ground transition, tracing  dense (e.g.,  for CO(5-4),  n$_{crit}=1.7 \times 10^5$ cm $^{-3} $ )  and thermally excited  gas in actively star-forming regions. 
 If many of those lines are observed, one could determine the thermal state of the gas and extrapolate down the observed CO excitation to infer the CO(1-0) excitation.  However, the full CO Spectral Line Energy Distribution (SLED)  is rarely observed: when it is available, it turns out that high-z DSGS  exhibit  a large diversity in the CO SLED  ranging from nearly thermalized through J=6, through subthermal  even at the J=3-2 line \cite{Casey2014}.  This is because CO has  a nonzero dipole moment (0.1 Debye), so that it will not reach equilibrium with the kinetic temperature of the gas via collisions, because spontaneous emission does have some effect instead   \cite{Goldsmith1972}, \cite{VanDishoeck1987}, \cite{Lyu1994}, \cite{Warin1996}.  It follows that there is no universal conversion from mid- and high-J CO lines to CO (1-0) for DSFGs:  converting down to CO(1-0) leads to significant dispersions, which make these lines unreliable  tracers of the cold molecular gas. Furthermore, the densities and temperatures required to excite these higher transitions may not be reached by the majority of the molecular gas, so that significant amounts of low-excitation gas may lurk in the DSFGs environments. Also, the non-collisional excitation of these lines prevents us from finding a clear connection with the dense gas, for which other tracing molecules are traditionally used.  
\noindent

 \section{\bf H$_2$O: the beacon of star formation} 
 \label{sec:H2O}
Water is expected to be one  of the most abundant molecules in molecular clouds,  after H$_2$ and CO.  In the coldest clouds, where H$_2$O freezes-out onto dust grain, its fractional abundance is  $< 10^{-8}$ , raising to $>10^{-4}$ in warm gas and shocks, due to ice evaporation or sputtering and exothermic gas phase reactions (see, e.g., \cite{Bergin1998}, \cite{Bergin2000},  \cite{Hollenbach2009}, \cite{Cernicharo2006}, \cite{Caselli2010}, \cite{Gonzalez2013}. \\
The water molecule  has a very large dipole moment of 1.84 D, which allows a strong coupling with the radiation field.  Furthermore, due to the high spacing between rotational levels (compared to other molecules with low-level transitions in the millimetric range), H$_2$O has a large number of rotational transitions, lying in the submm and FIR wavelength regime.
This combination of high dipole moment and peculiar rotational ladder makes water a very powerful tracer of  dense (n$_{\rm H}= 10 ^{5}-10^6$ cm$^{-3}$), warm  (T$_{\rm dust}\sim 50-100 $ K, T$_{\rm gas}=100-200$ K) star formation regions and, in general, of highly energetic processes, such as starbursts, shocks and AGN. This is not only because, in these conditions,  the dust temperature is raised above the ice evaporation temperature, boosting the H$_2$O abundance; but, most importantly,   the environmental  FIR radiation emitted by warm dust is able to populate high-J (E$_{\rm up}> 200$ K) levels of ortho- and para-H$_2$O  (FIR pumping, \cite{Takahashi1983}, \cite{Gonzalez2004}, \cite{Gonzalez2008}, \cite{Gonzalez2010},\cite{Weiss2010}, \cite{Gonzalez2014},\cite{Santaella2017},\cite{Gonzalez2022}), which then relax via cascade down to lower rotational levels emitting intense submm lines. \\
The high dipole moment of  H$_2$O  corresponds to large Einstein coefficients for spontaneous emission, which,  under the optically thin case, implies critical densities of the order of $10^8-10^9$  cm$^{-3}$ \cite{Poelman2007}. Radiative trapping does reduce the effective density for collisional excitation of the optically thick lines. For example, the warm gas model by \cite{Liu2017} shows that, for n$_{\rm H}$ = $10^{5}$ cm$^{-3}$, kinetic gas temperature T$_{\rm gas}$=50 K and fractional abundance X(H$_2$O)=$10^{-7}$ , the FIR pumping from dust continuum is driving the thermalization up to levels of increasing E$_J$ for increasing dust temperature. The low-excitation lines become weaker in the warm and hot regions where infrared pumping dominates over collisions. \\
As T$_{\rm dust}$ approaches T$_{\rm gas}$, the combination of collisions and pumping populates the levels in such a way that collisions drive the o-H$_2$O (p- H$_2$O) toward thermalization at T$_{\rm gas}$ for levels with E$_J \leq $ 200 (100) K: collisions still dominate over FIR pumping in populating the o-H$_2$O (p- H$_2$O) levels with E$_J \leq $ 350 (250) K, but the radiative pumping becomes the dominant source of excitation for the levels with E$_J \geq $ 350 (250) K. 
Viceversa, in the absence of FIR continuum, and under the same gas conditions, the o-H$_2$O (p- H$_2$O) populations could be excited by collisions only up to levels with energies up to 350K (250 K), corresponding to the o-4$_{14}$ (p-3$_{13}$) level. The  low-J lines can then be collisionally excited, and observed in emission,  also in regions where the other lines do not emit owing to weak far-IR continuum. \\
In general, the observed intensities of molecular emission depend on a complex competition between radiative and collisional processes. This is particularly true for the excitation of H$_2$0, where the FIR radiation from dust is strongly affecting the level populations, so that the interpretation of optically thick water emission lines requires a full radiative transfer calculation and an accurate modeling of the environment gas and dust mixture. 
High-z DSFGs have been investigated in their water emission lines, often detected in strongly lensed sources, by several works,   ( \cite{Omont2011}, \cite{Lis2011}, \cite{VanDerWerf2011},\cite{Jarugula2019}, \cite{Bradford2011}, \cite{Combes2012}, \cite{Lupu2012}, \cite{Bothwell2013}, \cite{Omont2013}, \cite{Vieira2013}, \cite{Weiss2013}, \cite{Rawle2014}, \cite{Yang2016}, \cite{Perrotta2023}),  which confirmed that those lines are amongst the brightest ones in the extreme star-forming environments of this galaxy population. Water lines have also been detected in absorption, and related to massive galactic outflows, possible quenching mechanism for star formation \cite{Jones2019}. \\
Being water a unique tracer of the FIR radiation field, it is not surprising that  a strong correlation has been found  (see \cite{Omont2013}
, \cite{Yang2013},\cite{Yang2016})  between the IR luminosity and the submm H$_2$O emission (with the exception of the ground-state emissions of para and ortho water, which, as we said, are largely affected by collisional excitation) . This dependence extends over 4 orders of magnitude of the luminosity range, regardless of the presence of a strong AGN signature. In particular, the sample of  strongly lensed high-z DSGFs presented in \cite{Omont2013} and \cite{Yang2013} and discovered in the Herschel Astrophysical Terahertz Large Area Survey   (H-ATLAS) shows this correlation f the p- H$_2$O  2$_{11}-2_{02}$ (${\nu} _{\rm rest}=$752.033 GHz, E$_{\rm up}$ = 137 K) and for the p- H$_2$O  2$_{02}-2_{11}$ (${\nu} _{\rm rest}=$ 987.927 GHz, E$_{\rm up}$ = 101 K), finding that, for these lines, L$_{\rm IR}$ varies approximately as   $\sim {L_{\rm H_2O}}^{1.2}$. As discussed in \cite{Perrotta2023}, this superlinear relation may be suggestive of the fact that these lines are still partially excited by collisions, so that they do not just trace the star formation through FIR pumping. Indeed, it was suggested in  \cite{Perrotta2023} that a linear, more direct correlation between the star formation process, as traced by the FIR luminosity, and the water excitation, could arise when considering those lines which are only associated to FIR pumping, i.e. the higher E$_{\rm up}$ lines. \\
We wish to emphasize that obtaining relatively high-resolution images of molecular emission from high-z DSFGs inevitably requires relying on the selection of strongly lensed sources. This introduces the additional complication of reconstructing the unlensed image of the galaxy.
Despite the difficulties in lens modeling, recent works by \cite{Giulietti2023} and \cite{Perrotta2023} could analyze high-resolution ($\lesssim 0.1-0.3$ arcsec) ALMA  
observations of the strongly lensed galaxy HATLASJ113526.2-01460, an optical/Near IR dark SB (with SFR of $\sim 900$ M$_{\odot}$ yr$^{-1}$) at redshift z $ \sim $3.1 discovered in the GAMA 12th field of the H-ATLAS survey. Accurate lens modeling and source morphology reconstruction in three different submillimeter continuum bands and in the C[II] , CO(8-7), and in three water emission lines (the low-J p-H$_2$O  $2_{02}-1_{11}$ line, the mid-J  o-H$_2$O  $3_{21}-3_{12}$ line, and the high-J p-H$_2$O  $4_{22}-4_{13}$  line) lead to the first high-resolution spatial map of the water emission in a high-z SB.  While the low-, medium- and high-J water emission lines appear to peak in a compact nucleus of  $\sim$ 500 kpc, where the FIR continuum also peaks, the image show a more extended molecular region where the p-H$_2$O  $2_{02}-1_{11}$ line is excited. The existence of plausibly cold  molecular gas, exhibiting water excitation at low energy levels at large distances  ($\sim$ 1 Kpc ) from the compact, star-forming core, surely deserves a deeper investigation. 

\section{Dense gas tracers}\label{sec:dense}
\cite{Bethermin2018}
The actual link between star formation and molecular gas resides in the high-density gas.  The first investigations of the role of dense  gas in molecular clouds was motivated by early Galactic observation, which found a correlation between the dense gas component and young stellar objects \cite{Evans1999}.  One key question is how much dense gas is needed to trigger star formation, and if the ratio between dense and diffuse molecular gas is correlated to the star formation efficiency \cite{Kauffmann2017}. \\
Direct tracers of dense gas include emission lines from molecules with high electric dipole, such as HCN, HNC, HCO$^{+}$ and CS, since the ground-state rotational transition of these molecules has a critical density for collisional excitations with H$_2$ of about $10^5$ cm$^{-3}$(see \cite{Gao2004},  \cite{Casey2014} and references therein). \\ 
Although CO is way more abundant that these molecules,  the low critical density of low-J CO transitions (n$_{crit} \sim 10^2$ cm$^{-3}$ ) makes CO(1-0) a reliable  tracer of  the bulk of molecular gas reservoir (diffuse molecular phase, with A$_V \sim 2 $, n$_H \sim 10^2$ cm$^{-3}$, T $\sim$ 100 K) rather than of the dense cores where stars form (n$_H$ $>10^5$ cm $^{-3}$  ). On the other hand, the mid and high-J CO emission lines are much affected by non-collisional excitation, preventing them to directly trace high-density environments  \cite{Greve2014}, \cite{Liu2015}, \cite{Kamenetzky2016} (see Sec. \ref{sec:h2co}).  \\
Assuming that HCN(1-0)/CO(1-0) measures  the dense gas fraction with respect to the diffuse gas, \cite{Gao2004} found that,  in nearby galaxies, the emission ratio  HCN(1-0)/CO(1-0) increases with the star formation rate.  Moving to DSFGs at z$\geq1$, it is not yet clear whether SFR is correlated with the dense  molecular gas fraction \cite{Oteo2017} or with the total molecular gas \cite{Rybak2022}, because of the sparse detections of HCN(1-0) in such galaxies: only three  DSFGs have been detected in HCN(1-0) to date (J1202 \cite{Rybak2022}, J16359 \cite{Gao2007} and SDP.9 \cite{Oteo2017}). The inferred HCN/CO ratio in these galaxies is not consistent with a universally high dense gas fraction in DSFGs \cite{Rybak2022}. \\
As for L$_{\rm HCN(1-0)}$/ L$_{\rm FIR }$ , assumed to be a  proxy for dense gas star formation efficiency,  there are indications for a sublinear relation at high redshifts \cite{Rybak2022}, which corroborates theoretical predictions by \cite{Krumholz2007} and \cite{Narayanan2008}.  However,  more statistics is needed to analyze the role of dense gas in the efficiency of star formation.  \\
It is important to notice that all these considerations rely on the assumption that the HCN (J = 1–0) transition is unambiguously associated with gas at H$_2$  densities  $ \gg 10^4$ cm$^{-3}$. If this is true, the mass of gas at densities  $ \gg 10^4$ cm$^{-3}$ could be inferred from the luminosity of this emission line. However, observations of the Orion A molecular cloud suggest that this line is actually associated to moderate gas densities, of about $ 10^3$ cm$^{-3}$ \cite{Kauffmann2017}, and that the only molecule really tracing dense gas is N$_2$H$^+$. In Orion A, the characteristic densities derived for the HCN (1–0) line are about two orders of magnitude below values commonly adopted in extragalactic environments, suggesting that only a fraction of the HCN (J = 1–0) luminosity traces the dense gas, while about half of the emission is related to the gas surrounding the densest clouds. 
On the other hand,  in high-z DSFGs  there is not sufficient angular resolution to analyze the single molecular clouds: once again, local observations and astrochemical modeling of the molecular  ISM have to be the primer for any attempt of high-z molecular spectra interpretation. \\
A recent deep spectral line survey, using NOEMA,  targeted the strongly lensed  DSFG NCv1.143 \cite{Yang2023},  and revealed an unprecedented large inventory of molecular species for a starburst at z=3.655.  Mid-J lines of HCN are detected, together with other dense gas tracers.  The analysis of the spectral lines is suggestive of a top-heavy stellar initial mass function and of high cosmic ray ionization rates.  However, we outline again that, in order to remove the degeneracy between molecular abundances and excitation of molecular lines, and to get the correct interpretation of the ISM physics, a solid chemical network including dust surface chemistry,  and high angular resolution observations, are needed. \\
\subsection{AGN tracers}
Dense gas  acts as a reservoir for sites of massive star formation, but it is also a  fuel for active galactic nuclei (AGN). In the framework of galaxy evolution, the study of high-redshift DSFGs  is of paramount importance to address the issue of coevolution between galaxies and supermassive BHs \cite{Alexander2012}, \cite{Kormendy2013}, \cite{Mancuso2017}. 
As heating sources, AGNs can radiatively and mechanically alter the chemical composition and the excitation state of the surrounding molecular medium. It has been proposed that different heating mechanisms produce specific, distinguishable signatures in the surrounding interstellar medium. Specifically, the main powering source in starburst comes from nuclear fusion, with the intense UV flux producing photodissociation regions (PDRs) around massive stars,  while, in the vicinity of an AGN, the strong X-ray emission produces, thanks to the higher penetrating capabilities of this radiation, X-ray dominated regions (XDRs) larger in volume than PDRs. Mechanical heating due to AGN outflows or to supernovae, as well as cosmic rays, is also shaping the chemical composition and the molecular excitation state.  This should lead to distinguishable  properties of the dust and molecules surrounding the gas. \\
One key problem lies the high resolution needed to probe the very central region of AGNs: the kinematics of the gas flow in the central $\lesssim$ 100 pc is a crucial information to account for the different physical conditions \cite{Fathi2013}. This makes necessary to rely on  millimeter/submmillimeter interferometric spectroscopic observations, because of their high spatial and spectral resolution. Also, such wavelenghts are not affected by dust extinction, which is of paramount importance if the target AGN is enshrouded in dust. \\
Following this idea, it has been proposed to use the line ratios of specific transitions of dense gas-tracing molecules as a possible diagnostic to discriminate the main heating source in  galactic nuclei. 
An enhanced intensity of HCN(1-0) compared  to HCO$^{+}$(1-0) has been proposed as a unique feature to AGNs (\cite{Jackson1993},\cite{Tacconi1994}, \cite{Sternberg1994}, \cite{Kohno2001},\cite{Imanishi2007}, \cite{Krips2008}, \cite{Davies2012}, \cite{Aalto2015}, \cite{Imanishi2016}, \cite{Oteo2017}). However, low values of  HCN(1-0)$ / $HCO$^{+}$(1-0) (\cite{Sani2012}) have been detected in AGNs, as well as high values in non-AGNs (\cite{Costagliola2011}, \cite{Snell2011}). This diagnostic has been later questioned by \cite{Privon2015}, showing that the ratio of the HCN(1-0) and HCO$^{+}$ (1-0) integrated intensity emission cannot be simply interpreted in terms of the AGN or starburst  dominance, because it is likely affected by multiple processes, including  contamination from a coexisting starburst, effects of density, opacities, temperature, radiative effects (non- collisional excitation), outflowing material and abundances. \\
  Following the indication by \cite{Kauffmann2017}  that HCN and HCO$^+$  J=1 $\rightarrow$0 emission could be dominated by low gas densities, \cite{Zhou2022} argued that a more suitable star formation tracer is the J=2 $\rightarrow$1 transition, having higher critical density ($1.6 \times 10^6 $ and $2.8 \times 10^5$ cm$^{-3}$, respectively, for HCN and HCO$^+$  J=2 $\rightarrow$1 line). However, no significant difference was found between the average HCN/HCO$^+$ ratio in the a sample of eight AGN-dominated galaxies and eleven nearby star-formation dominated galaxies. \\
In this view, \cite{Imanishi2010}, \cite{Izumi2013},\cite{Izumi2015}, \cite{Izumi2016} examined higher rotational lines of dense-gas tracing molecules, since higher resolution is achievable for these lines compared to the fundamental (1-0) transition. Specifically, the integrated line intensity ratio HCN(4-3)/HCO$^{+}$(4-3) and HCN(4-3)/CS(7-6) seems to be enhanced in AGN when compared to pure Starburst ({\it submillimeter HCN enhancement}). The physical interpretation is not unique, though: a chemical layout focusing on high-temperature chemistry was invoked by, e.g., \cite{Izumi2013} and by \cite{Harada2021}, to explain the HCN/ HCO$^{+}$ enhancement through an astrochemical mechanism that boosts the HCN abundance relative to HCO$^+$.\\
The neutral-neutral reactions: 
O+H$_2$ $\rightarrow$ OH+H  \ \ \ and \ \ \ OH+H$_2$ $\rightarrow$ H$_2$O+H  
are efficiently activated at high temperature environments, especially at T$>300$ K, indicating that much of the elemental Oxygen is in the form of water. The Hydrogenation of CN reacting with molecular Hydrogen to produce HCN is a slightly endothermic reaction, which enhances the HCN abundances at high temperatures, whatever the heating source: 
\\ CN+H$_2$ $\rightarrow$ HCN+H.  
The HCO$^+$ ion is generally created by the reaction 
CO+H${_3}^+$ $\rightarrow$ HCO$^+$+H$_2$ . 
Although this ion tends to dissociatively  protonate a water molecule, producing a CO molecule, the abundance of HCO$^+$ with respect to H$_2$ at high temperatures remains approximately constant. Thus, once water formation is promoted by the high temperatures, the water-induced reactions facilitate HCN enhancement with respect to  HCO$^+$. This could explain why the observed abundance ratio of HCN-to-HCO$^+$ in the nucleus of AGNs seems to be enhanced by a factor of a few to even $\gtrapprox$ 10. However, X-rays are not the only possible heating mechanism, as  mechanical heating due to an AGN jet can contribute significantly . \\

\section{Summary and concluding remarks}\label{sec:summary}
We discussed the wealth and robustness of the informations provided by the current molecular lines observations in high-z DSFGs.  What arise from this review is that there are still many open questions that molecular lines data  could not clarify yet.   The main problem in the interpretation of spectroscopic lines from high-z sources resides in the fact that, even  for gravitationally lensed sources, there is a limit in the spatial resolution that prevents us from the characterization of single molecular clouds. So, we receive integrated signals, emitted by gas in different physical conditions. For this reason, we focus on those molecular lines which better trace specific densities and temperatures of the gas. \\
To properly interpret molecular line observations, the modeled intrinsic line emission has to be implemented in the radiative transfer equation which, in turn, encodes the ISM properties along the line of sight.   As a matter of fact, the ISM in high-z DSFGs differ from that of local or more quiescent galaxies on several respects.  At high redshifts, the CMB radiation can affect the observability of molecular lines, as well as the population of excited rotational levels. The high rates of star formation typical of DSFGs  boost the flux of FUV radiation, which   regulates the physics and chemistry of PDR regions. The strong interplay between the FUV radiation and the large amount of dust, also typical of this population, allows an efficient shielding against ionizing radiation for large part of the gas, which can turn from atomic to predominantly molecular. In dark molecular clouds,  cradle of simple molecules as well as complex organic molecules, a key role is played by cosmic rays: at the typical SFRs of this galaxy population, their flux is expected to be enhanced compared to quiescent galaxies, as observations seem to indicate. Cosmic rays deeply affect molecular chemistry, starting from the production of the pivotal ion ${\rm H^{+}}_3$, whose abundance drives all the chain of chemical reaction network to the equilibrium molecular abundances. Notably, cosmic rays will impact the photodesorption in dust grains, so that gas chemistry will be deeply linked to surface chemistry even in a starless molecular core.  \\
In summary, the unique and extreme ISM of DSFGs not only influences the radiative transport of emission lines and their observed brightness but also significantly impacts underlying chemistry and molecular abundances. 
We have then to face a problem of degeneracy between the radiative transport process and the molecular abundances: for example, unless a line is optically thin, a single bright emission line may indicate either a high excitation of the corresponding upper energy level, or just an increase of the molecular abundance for that particular species. In order to disentangle the two effects, it is crucial to have large spectral coverage and high spectral resolution observations, and to carefully insert all the “ingredients” in the astrochemical networks when implementing simulations of molecular clouds. This problem is exacebated  by lack of spatial resolution, which forces us to interpret the integrated emission lines in a statistical approach, rather than indicators or thermometers of a single molecular cloud.  \\
For all these reasons, caution needs to be taken when using an H$_2$-to-CO conversion factor to estimate the molecular content of a whole galaxy,  or when inferring global properties of the unresolved source from tracers molecules which are affected by different local environments through the ISM. \\
In view of future higher-resolution spectral observations, a statistical approach will be needed, in which the filling factors of the several phases of the ISM are estimated and used as inputs for astrochemical and radiative transfer models, to allow for an improved interpretation of the molecular lines and of the wealth of information they can potentially deliver. 


\vspace{6pt} 


\acknowledgments{This work is partially funded from the projects: ``Data Science methods for MultiMessenger Astrophysics \& Multi-Survey Cosmology'' funded by the Italian Ministry of University and Research, Programmazione triennale 2021/2023 (DM n.2503 dd. 9 December 2019), Programma Congiunto Scuole; Italian Research Center on High Performance Computing Big Data and Quantum Computing (ICSC), project funded by European Union - NextGenerationEU - and National Recovery and Resilience Plan (NRRP) - Mission 4 Component 2 within the activities of Spoke 3 (Astrophysics and Cosmos Observations); PRIN MUR 2022 project n. 20224JR28W "Charting unexplored avenues in Dark Matter"; INAF Large Grant 2022 funding scheme with the project "MeerKAT and LOFAR Team up: a Unique Radio Window on Galaxy/AGN co-Evolution; INAF GO-GTO Normal 2023 funding scheme with the project "Serendipitous H-ATLAS-fields Observations of Radio Extragalactic Sources (SHORES)”. F.P. wishes to acknowledge L. Boco and S. Viti for useful discussions. }

\conflictsofinterest{The authors declare no conflict of interest.} 

\appendixtitles{no} 
\appendixstart
\appendix


\begin{adjustwidth}{-\extralength}{0cm}

\reftitle{References}


\bibliography{main}
\end{adjustwidth}
\PublishersNote{}
\end{document}